\documentclass[manuscript]{aastex}

\shorttitle{A parsec-scale jet from Sgr A*}
\shortauthors{Li et al.}

\begin{document}


\title{Evidence for A Parsec-scale Jet from The Galactic Center Black Hole: Interaction with Local Gas}


\author{Zhiyuan Li, Mark R. Morris}
\affil{Department of Physics and Astronomy, University of California, Los Angeles, CA 90095}
\email{zyli@astro.ucla.edu; morris@astro.ucla.edu}

\and 

\author{Frederick K. Baganoff}
\affil{Kavli Institute for Astrophysics and Space Research, Massachusetts Institute of Technology, Cambridge, MA 02139}
\email{fkb@space.mit.edu}




\begin{abstract}
Despite strong physical reasons that they should exist and decades of search, jets from the Galactic Center Black Hole, Sgr A*, have not yet been convincingly detected.
Based on high-resolution {\it Very Large Array} images and ultra-deep imaging-spectroscopic data produced by the {\it Chandra X-ray Observatory},
we report new evidence for the existence of a parsec-scale jet from Sgr A*, by associating 
 a linear feature G359.944-0.052, previously identified in X-ray images of the Galactic Center,
 with a radio shock front on the Eastern Arm of the Sgr A West HII region.
We show that the shock front can be explained in terms of the impact of a jet having a sharp momentum peak along the Galaxy's rotation axis,
whereas G359.944-0.052, a quasi-steady feature with a power-law spectrum, can be understood as
synchrotron radiation from shock-induced ultrarelativistic electrons cooling in a finite post-shock region downstream along the jet path.
Several interesting implications of the jet properties are discussed.
\end{abstract}





\section{Introduction} {\label{sec:intro}}
Highly collimated, magnetized outflows of relativistic particles (i.e., {\it jets}) emanating from galactic nuclei are generally thought to be powered by the ``central engine'', i.e.,
accretion onto a super-massive black hole (SMBH). Studies of jet-related phenomena hold promise for advancing our knowledge on not only the physics of accretion, but also the interaction 
and co-evolution between the SMBH and its environment \citep[e.g.,][]{Rich98,McNa07,Fabi12}. 
This physical picture, however, remains rather elusive in the context of our own Galaxy, which hosts the nearest SMBH, commonly known as Sgr A* \citep{Meli01}.
Being exceptionally underluminous and variable due to still poorly understood physical processes \citep[see recent reviews by][]{Genz10,Yuan11,Morr12}, Sgr A* exhibits 
centimeter-to-millimeter emission whose broad properties (spectral, temporal and polarization) are suggestive of, and thus extensively modeled by,
synchrotron radiation from jets \citep[e.g.,][]{Falc93,Falc00,Mark01,Yuan02,Yuse06,Yuan09} that are probably
coupled with a radiatively inefficient, advection-dominated accretion 
flow \citep[cf.][]{Nara08}. However, despite its virtue of proximity,
Sgr A* is persistently seen as a compact radio source under the currently best available resolutions afforded by Very-Long-Baseline Interferometry (VLBI) observations, down to the 
vicinity of its presumed {\sl event horizon} \citep{Bowe04,Shen05,Doel08},
thus leaving room for doubt on the existence of the putative jet.
On the other hand, belief in the jet scenario has been reinforced by viewing Sgr A* as a scaled-down version of M81*, the SMBH in the nearby massive spiral galaxy M81. 
The latter exhibits a radio core having properties very similar to those of Sgr A* \citep{Reut96,Brun06}, whereas VLBI observations resolve a jet 
within 100-1000 Schwarzschild radii on one side of the core \citep{Biet00,Ros12}.
In parallel, recent theoretical effort \citep{Mark07} has attempted to reconcile the apparent size of Sgr A* and the hypothetical jet, deriving
constraints on its inclination and position angles.

Regardless of the correctness of the jet scenario for the compact source itself, valuable insights on the interplay between the SMBH and its environment 
can be provided by possible jet/outflow manifestations beyond the immediate vicinity of Sgr A*. 
In the past decades at least seven suggestions of this kind have been made at various physical scales and wavelengths: (i) A narrow ridge of low-frequency radio emission extends out to
$\sim$30 pc from the Sgr A radio complex, roughly following the Galaxy's rotation axis toward negative latitudes \citep{Yuse86};
(ii) A radio filament, sometimes called the Galactic Center Spur, appears to protrude from the Galactic Center (GC) into the northern Galactic bulge with an extent of $\sim$3 kpc \citep{Sofu89}; 
(iii) Within the inner half-parsec of the GC, a $\sim$0.1 pc-long, slightly-curved near-infrared (NIR) filament pointing to the position of Sgr A* to within $\sim$10$^\circ$ \citep{Ecka06};
(iv) A bipolar outflow oriented roughly parallel to the Galactic plane, inferred from the bow-shock shape of two clumps and the morphologies of several 
filaments seen in the NIR within the inner 0.2-pc of the GC (\citealt{Muz07,Muz10}; also mentioned by \citealt{Ecka06});
(v) An X-ray filament, designated {\sl G359.944-0.052}, lies within the central parsec of the GC (\citealt{Muno08}; see also \citealt{Morr04}). The sharp and straight appearance of this feature allows for a well-defined orientation that, intriguingly, points to the position of Sgr A* and is coincident with the Galaxy's rotation axis in projection;
(vi) A pair of linear gamma-ray features extending up to $\sim$8 kpc from the GC in opposite directions
at an angle of $\sim$15$^\circ$ from the Galaxy's rotation axis \citep{Su12};
(vii) Within the inner few parsecs of the GC, a linear radio feature and several radio blobs, which are roughly colinear with Sgr A*, are interpreted as manifestations of a jet 
oriented at a position angle of $\sim$$60^\circ$ ($\sim$65$^\circ$ relative to the Galaxy's rotation axis; \citealt{Yuse12}).

In all the above cases, the indication for a collimated outflow from Sgr A* is essentially morphological, that is, a causal relation with the SMBH is argued on the basis of positional coincidence,
albeit in some of these cases massive young stars in the GC might equally be considered as the cause \citep[e.g.,][]{Muz07,Carr13}.
When put together, the orientations of these candidates tend to contradict rather than corroborate each other for a coherent jet/outflow\footnote{We point out a few correspondences that could link the 
various suggestions:  Candidates (ii) and (vi) -- a visual examination (e.g., see Fig.~2 in \citealt{Carr13}) shows that 
the Galactic Center Spur (Sofue et al.~1989) and the northern segment of the gamma-ray ``jet'' \citep{Su12} are largely spatially coincident.  Their curved morphologies are not obviously consistent with a collimated jet, and both features have been interpreted as part of a biconical outflow driven by recent star-forming
activities in the GC \citep{Carr13}; Candidates (iv) and (vii) -- their orientations are broadly consistent with each other, i.e., roughly along the Galactic plane. Its remains plausible  
that they trace the same outflow produced either by Sgr A* or the central star cluster \citep{Ward92, Muz07}; Candidates (i) and (v) -- these are co-aligned, but are present on quite different scales, as discussed below in \S~\ref{sec:discussion}; 
Candidate (iii) has a similar position angle as candidate (v), but ``the fact that it is bent and not pointing directly towards SgrA* may question'' it being a jet \citep{Ecka06}.}.
Moreover, neither direct measurement of the kinematics nor tight constraint on the energetics of any of these candidates is currently available, making it difficult to confront them with the expected
jet kinematics and jet power. Thus, none of the suggested cases has yet been widely accepted as convincing imaging of the long-sought jet from Sgr A*, and they all invite further examination.

Motivated by recent multi-wavelength, high-resolution observations of the GC, in particular those obtained by the Very Large Array (VLA; \S~\ref{sec:data}), here we revisit one of the jet candidates, the linear X-ray feature G359.944-0.052, 
and provide synthesis evidence that it is indeed associated with a parsec-scale jet from Sgr A* interacting with the local gas (\S~\ref{sec:result}). We conclude our study with
several interesting implications of the jet properties (\S~\ref{sec:discussion}).

\section{Data preparation} {\label{sec:data}}
We collect VLA 1.3\,cm and 3.6\,cm images of the GC, originally presented by \citet{Zhao09}
in their study of the dynamics and geometry of the well-known Sgr A West HII region \citep[cf.][]{Morr96}. In particular, the 1.3\,cm image offers one of the highest available angular resolutions
(full width at half maximum $\approx$ $0\farcs2{\times}0\farcs1$) and thus promises to reveal interesting substructures of the ionized gas in Sgr A West.

We utilize {\sl Chandra} X-ray observations of the GC taken between September 1999 and March 2011, which have accumulated a total exposure of 1.46 Ms and provide 
an unprecedented temporal baseline nearly doubling that achieved in \citet{Muno08}.
In all these observations, the aim-point was placed on the I3-chip of the Advanced CCD Imaging Spectrometer (ACIS) and was within 30$^{\prime\prime}$ 
of Sgr A$^\ast$, ensuring an optimal angular resolution of $\lesssim$ 0\farcs4.
We have downloaded from the public archive the corresponding data and uniformly reprocessed them following the standard procedure, using
CIAO v4.4 and the corresponding calibration files\footnote{http://cxc.harvard.edu/ciao/}. The relative astrometry among individual observations is calibrated by matching centroids of 
point-like sources detected within the common field-of-view. We restrict our analysis to the 2-8 keV energy range, with the lower and higher limits set by the heavy 
foreground absorption and the instrument sensitivity, respectively. 

Finally, we make use of the [Ne II]$\lambda$12.8\,$\mu$m data cube of Sgr A West \citep{Iron12}, which was obtained by the Texas Echelon Cross Echelle Spectrograph (TEXES) on the NASA Infrared Telescope 
Facility. The data cube has a spectral resolution of 4 km~s$^{-1}$ and an angular resolution of $\sim$1\farcs3.





\section{The multi-wavelength imprint of a parsec-scale jet} 
\begin{figure}
\epsscale{1.0}
\plotone{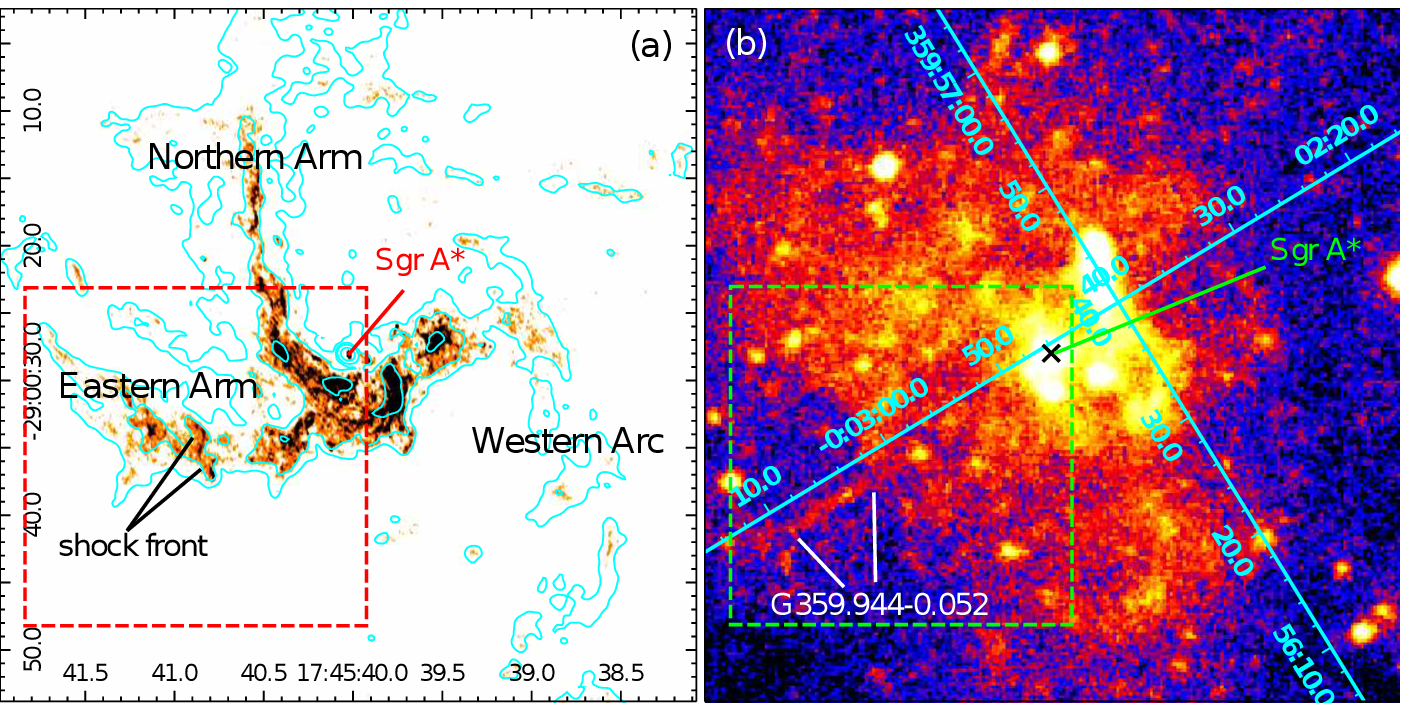}
\caption{The inner 2\,pc$\times$2\,pc region centered on Sgr A*, the SMBH in our Galaxy. (a): A VLA 1.3\,cm continuum image
tracing the three arms of ionized gas in Sgr A West. A ``$\langle$''-shaped feature on the Eastern Arm, suggestive of a shock front, is marked.
VLA 3.6\,cm intensity contours are overlaid. (b): A {\sl Chandra} 2-8 keV image of the same region, highlighting
the linear feature G359.944-0.052, which points to the location of Sgr A* following the Galaxy's rotation axis. J2000 celestial and Galactic coordinates are shown
in (a) and (b), respectively. The dashed box delineates a sub-region further shown in Fig.~2. \label{fig:jet}}
\end{figure}

\begin{figure}
\plotone{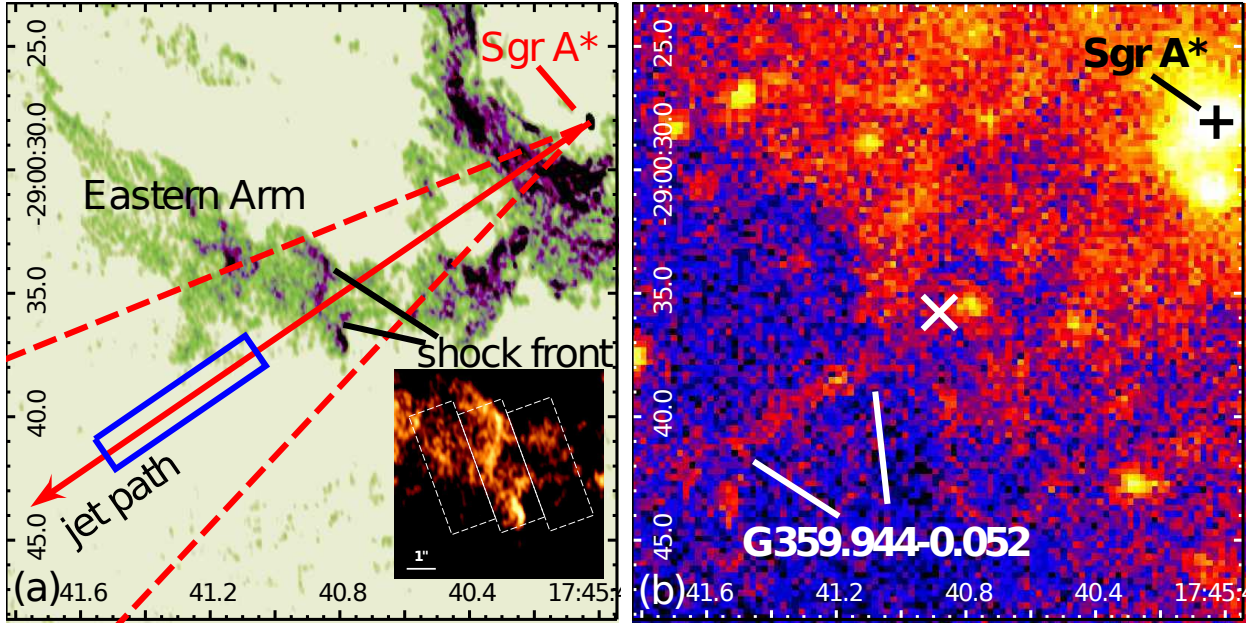}
\caption{A close-up view of the shock front and G359.944-0.052. (a) The Eastern Arm in VLA 1.3\,cm continuum, highlighting the ``$\langle$''-shaped shock front. 
An imaginary line connecting Sgr A$^\ast$ and the apex of the ``$\langle$''-shape {\sl naturally} passes through the long-axis
of the X-ray feature G359.944-0.052 (represented by the blue rectangle), suggesting the presence of a jet that
creates both the shock front and the X-ray feature. A pair of red dashed lines, defining an opening angle of $25^\circ$, outline a possible cocoon of the jet, which
we suggest is responsible for the shape and extent of the shock front. The three rectangles in the inset
mark the regions from which the [Ne II]  flux-velocity diagrams (Fig.~\ref{fig:neii}) are constructed.
(b) The {\it Chandra} 2-8 keV view of the same region, highlighting G359.944-0.052.
The apex of the radio shock front, the assumed primary site of particle acceleration, is marked by an ``X''. J2000 celestial coordinates are shown.
\label{fig:zoom}}
\end{figure}

{\label{sec:result}}
Fig.~\ref{fig:jet} shows the inner 2\,pc$\times$2\,pc region of the GC at radio and X-ray wavelengths. Throughout this work a distance of 8 kpc is adopted for the GC (1$^{\prime\prime}$ corresponds 
to $\sim$0.04 pc; \citealt{Ghez08,Gill09}).
While the X-ray image (Fig.~\ref{fig:jet}b) exhibits a mixture of point-like sources and diffuse features \citep{Baga03}, including the linear feature
G359.944-0.052, most prominent in the radio image (Fig.~\ref{fig:jet}a) is Sgr A West, with its well-known three-armed morphology.
Here we focus on a narrow ($\lesssim$ 0\farcs3) feature lying across the short-axis of the Eastern Arm,
which consists of two joined arc segments together forming a ``$\langle$''-shape (see Fig.~\ref{fig:zoom}a for a close-up view). The upper (northern) arc is the sharper and more extended of the two,
with its upper end bending along the edge of the Eastern Arm, whereas the lower (southern) arc appears less smooth and brighter at its lower half.
We note that this feature is also evident on various other tracers of Sgr A West, e.g.,
3.6\,cm continuum (shown as contours in Fig.~\ref{fig:jet}a), H92$\alpha$ emission \citep{Robe93}, [Ne II]$\lambda$12.8\,$\mu$m emission (Irons et al.~2012), and Paschen-$\alpha$ emission \citep{Scov03,Wang10}.
Best resolved by the VLA 1.3\,cm image, this remarkable feature, which is unique within Sgr A West in shape and sharpness, strongly suggests the presence of a shock front. 

\begin{figure}
\centerline{
\includegraphics[angle=0,scale=0.6]{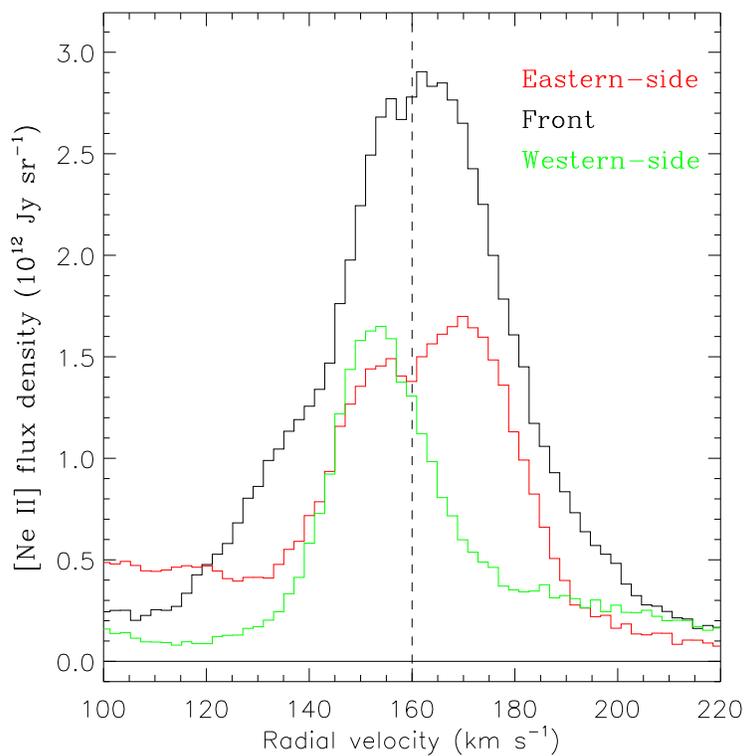}
}
\caption{[Ne II] flux density-velocity diagrams for the front ({\it black}), its eastern side ({\it red}) and western side ({\it green}) along the Eastern Arm. Compared to the front
and the eastern side, the western side exhibits a significant flux drop at velocities above $\sim$160 km~s$^{-1}$, indicating a momentum impact on the otherwise continuous
velocity field of the Eastern Arm.}
\label{fig:neii}
\end{figure}

To examine this possibility, we construct from the TEXES data [Ne II] flux density-radial velocity diagrams for the ``front'' and its immediate vicinity along the Eastern Arm (see the inset of Fig.~\ref{fig:zoom}a). 
As shown in Fig.~\ref{fig:neii}, these diagrams reveal a significant depression in the [Ne II] emission at velocities between 160 and 190 km~s$^{-1}$, and hence a relative paucity of gas, on the immediate western side of the front, as compared 
to that on the eastern side (see also the radio continuum image in Fig.~\ref{fig:zoom}a). This depression is mostly due to missing velocity components (above $\sim$160 km~s$^{-1}$) that are present
both in the front and on its eastern side. Regarding the streaming motion of the Eastern Arm, such that gas flows in from the east \citep[e.g.,][]{Zhao09}, 
the strong velocity discontinuity and the associated gas depletion to the west of the front can be naturally attributed to the impact of 
an external source of momentum arriving from the west at a supersonic velocity, i.e., a shock.
Hereafter we shall refer to this remarkable feature of ionized gas as the ``shock front''\footnote{The morphology of this feature strongly suggests to us that it be designated the ``Seagull Nebula'' for future reference.}. 

Strikingly, a spatial relation appears to exist among Sgr A*, the shock front and G359.944-0.052, in the sense that an imaginary line connecting
Sgr A* and the apex of the ``$\langle$''-shape (at an offset of [$\Delta\alpha$, $\Delta\delta$] $\approx$ [12\farcs8, -7\farcs7] from Sgr A*) {\sl naturally} passes through the long-axis of
the X-ray feature\footnote{We have registered the X-ray and radio images according to the centroid position of Sgr A*, the accuracy of which is better than $\sim$0\farcs1.} (Fig.~\ref{fig:zoom}a).  We suggest that such a well-constrained geometry is not a mere coincidence. {\sl Rather,
this is precisely what would be expected from the shock interaction of a jet from Sgr A* with the inflowing gas in the Eastern Arm and, as a direct consequence,
X-rays arising from the post-shock region downstream along the jet path}.

\begin{figure}
\epsscale{0.7}
\plotone{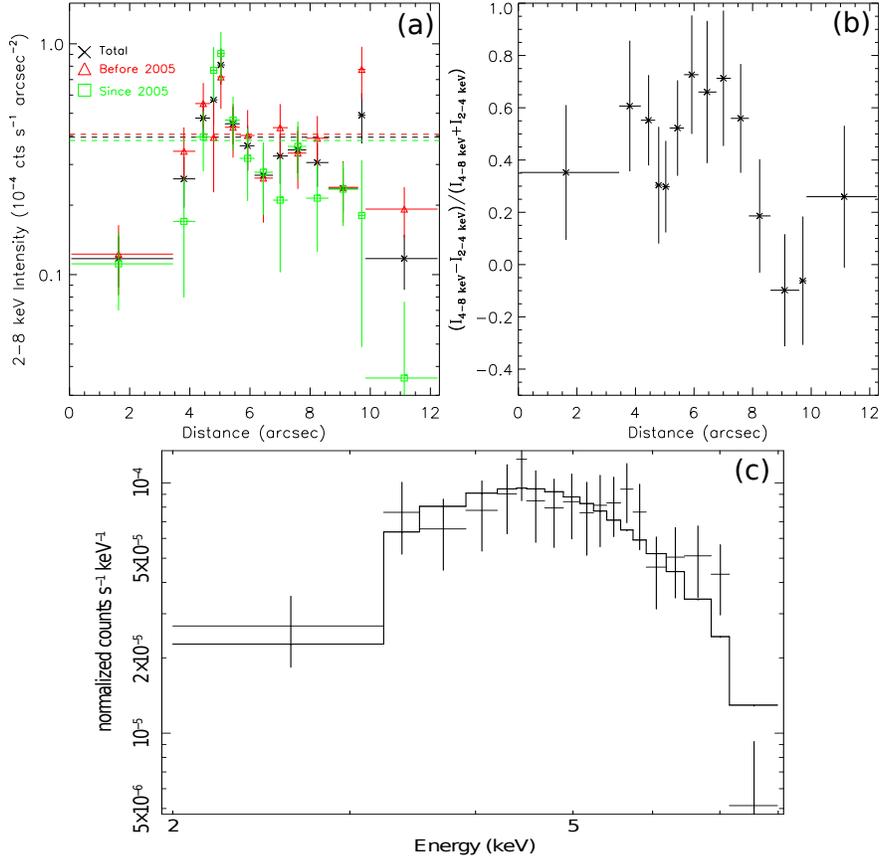}
\caption{Observed X-ray properties of G359.944-0.052. (a) Intensity profiles of G359.944-0.052 as a function of distance from
the apex of the shock front along a position angle of 124\fdg5.
Different color symbols represent profiles extracted from the total exposure ({\sl crosses}) and two epochs before ({\sl triangles}) and since ({\sl squares}) 2005.
Spatial binning is adaptive to achieve a minimum of 40 counts and a S/N better than 4 in the {\sl total} profile.
The levels of the subtracted background are indicated by the horizontal dashed lines.
There is no significant variation between the two epochs. (b) The profile of hardness ratio between the 4-8 keV and 2-4 keV bands.
A spectral softening at the far-side of G359.944-0.052 is evident.  
(c) The average spectrum of G359.944-0.052, 
adaptively binned to achieve a minimum of 30 counts and a S/N better than 3.
The histogram represents the best-fit absorbed power-law model, including dust scattering. The fit is acceptable with $\chi^2/d.o.f. = 11.6/16$.}
\label{fig:xray}
\end{figure}

We investigate this hypothesis by examining the X-ray properties of G359.944-0.052 with the {\sl Chandra}/ACIS data. As shown in Figs.~\ref{fig:jet}b and \ref{fig:zoom}b, 
the linear feature G359.944-0.052 clearly stands out from the surrounding diffuse emission, and conspicuously points to the position of Sgr A* along an orientation coincident with the Galaxy's rotation axis.
We extract an intensity profile for G359.944-0.052 along its long-axis, i.e., the hypothetical jet path (at a position angle of 124{\fdg}5$\pm$1\fdg5, east of north; Fig.~\ref{fig:zoom}a),
averaged over a full width of 1\farcs5 in the perpendicular direction.
We have verified that the feature is unresolved along its short-axis, by comparing it with nearby point-like sources. 
A local background is adopted from adjacent regions running parallel to G359.944-0.052. As shown in Fig.~\ref{fig:xray}a, the linear feature becomes prominent at a distance of
$\sim$$3\farcs5$ from the apex of the shock front and gradually fades off by a distance of $\sim$$11^{\prime\prime}$ ($\sim$$25^{\prime\prime}$ from Sgr A*). An unresolved knot is also apparent
at a distance of $\sim$$5^{\prime\prime}$, although we cannot rule out the possibility of a superposition of an unrelated source belonging to one of the least luminous classes of 
X-ray sources found in the GC \citep{Muno03}. 
We note that X-ray emission can in principle be present from further upstream along the jet path, in particular starting at the shock front,
but the relatively strong and non-uniform diffuse emission there causes unavoidable confusion. Hence we conservatively adopt a $7\farcs5\times$1\farcs5 rectangular 
region\footnote{Indeed this is different from \citet{Muno08}, who, based on poorer statistics, traced the feature up to a point-like source $\sim$1\farcs3 west of the 
apex (Fig.~2b), which we consider probably unrelated.} 
to define G359.944-0.052 (Fig.~\ref{fig:zoom}a), and find $\sim$458 net counts out of a total (source plus background) of 1058 counts, giving a $\sim$14$\,\sigma$ significance to G359.944-0.052.
We further probe temporal variation of the intensity profile by dividing the total dataset into two epochs of comparable exposure: one consisting of observations taken before 2005 and the other since 2005.
No statistically significant variation is found (Fig.~\ref{fig:xray}a), indicating that G359.944-0.052 has remained a quasi-steady feature on a 10-yr timescale.
On the other hand, the hardness ratio distribution along the feature (Fig.~\ref{fig:xray}b) reveals a spectral softening beyond a distance of $\sim$$8^{\prime\prime}$ from the shock front, which
is not due to any gradient in the foreground absorption \citep{Scho10}, but is rather consistent with progressive radiative cooling of the emitting particles in the hypothetical jet.

We have also extracted the spectrum of G359.944-0.052, averaging it over all observations.
A background spectrum is extracted, again from regions immediately adjacent to the feature. 
The spectrum appears featureless (Fig.~\ref{fig:xray}c) and hence we fit it with an absorbed power-law model, taking into account the effect of spectral hardening by 
foreground dust scattering with an $E^{-2}$ dependence \citep{Pred95}.
The fitted parameters are the foreground absorption column density \citep{Morr83}, $N_{\rm H}=11.6^{+5.2}_{-3.9}\times10^{22}{\rm~cm^{-2}}$ and the photon-index, 
$\Gamma=1.77^{+0.89}_{-0.76}$ (at 1\,$\sigma$ confidence level). The best-fit model predicts an intrinsic 2-10 keV luminosity of $2.4\times10^{32}{\rm~erg~s^{-1}}$. 
We have also fitted the spectrum with a thermal plasma model (APEC in XSPEC), but found an unphysically high temperature $\gtrsim$ 10 keV.

\begin{figure}
\epsscale{1}
\plotone{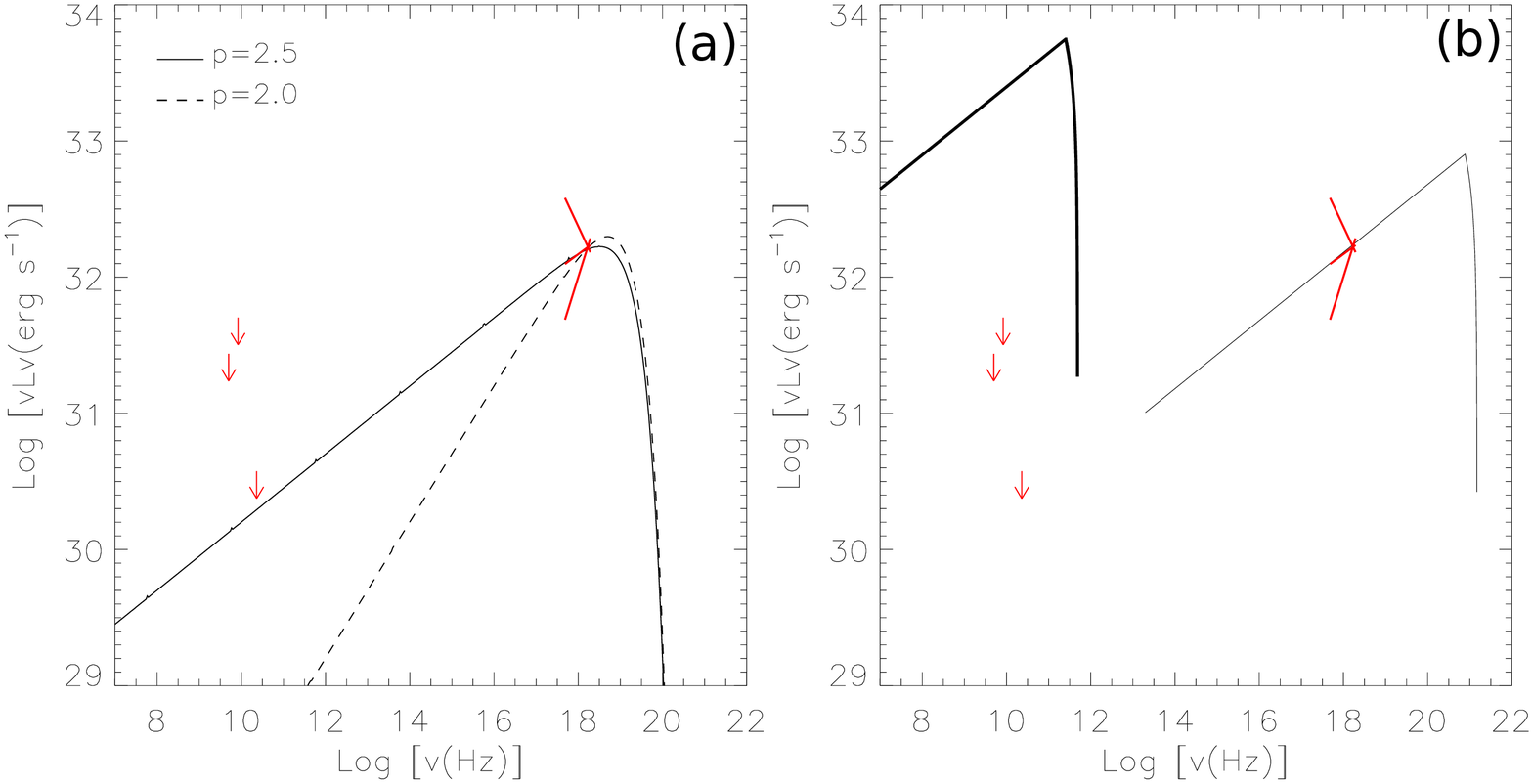}
\caption{(a) One-zone steady-state synchrotron models characterizing the observed X-ray power-law spectrum of G359.944-0.052, which are also consistent with
current 3\,$\sigma$ upper limits at radio frequencies (1.33, 49.8 and 45.0 mJy at 23, 8.4 and 5.0 GHz, respectively; \citealt{Muno08,Zhao09}), as indicated by the arrows. The solid and dashed curves represent 
models with different values of $p$, the power-law slope of the electron energy distribution.
(b) A steady-state model in which seed photons from within the GC scattering off the relativistic electrons
in the jet account for the observed X-ray spectrum (thin solid curve). Direct synchrotron radiation (thick solid curve) from the relativistic
electrons, however, significantly exceeds the radio upper limits. See text for details.}
\label{fig:model}
\end{figure}

The X-ray spectrum of G359.944-0.052
suggests a non-thermal origin (e.g., produced by synchrotron or inverse Compton radiation), as commonly encountered in extragalactic jets \citep[e.g.,][]{Harr02}.
To further test this possibility, we have created a simple one-zone steady-state model in which synchrotron radiation of
relativistic electrons with a canonical power-law energy distribution accounts for the X-rays.
With reasonable input parameters such as the magnetic field strength ($B$ = 1 mG),
a maximum Lorentz factor ($\gamma_{\rm max} = 10^8$), the bulk velocity of the jet (0.9 $c$) and a viewing angle of 90$^\circ$ (see below),
we find that the model\footnote{Calculations are assisted by an IDL program, called {\it Compton toys}, publicly available at http://www.jca.umbc.edu/{\textasciitilde}markos/cs/} (especially 
the $p=2.5$ case, where $p$ is the power-law slope of the electron energy distribution, $N(\gamma) \propto \gamma^{-p}$)
can reproduce the slope of the X-ray spectrum and in the meantime be consistent with the current non-detection of radio counterparts for G359.944-0.052 (Fig.~\ref{fig:model}a).
Furthermore, the model indicates a synchrotron cooling timescale of $1.2(B/1{\rm~mG})^{-1.5}(E/5{\rm~keV})^{-0.5}$ yr, which agrees well with the apparent length of the X-ray feature ($\sim$0.4 pc if measured from the shock front).

We further examine models involving inverse Compton (IC) radiation, adopting similar magnetic field strength ($B$ = 1 mG),
bulk velocity of the jet (0.9 $c$), and power-law slope ($p=2.5$) as in the favored synchrotron model, but also a maximum Lorentz factor of $2\times10^4$. We assume that the seed photons
are contributed externally and predominantly by reprocessed dust emission from within the GC, with an energy density of $\sim$$3\times10^{-8}{\rm~erg~cm^{-3}}$
at a peak frequency of $6\times10^{12}{\rm~Hz}$ \citep{Davi92}. We find that the predicted synchrotron radiation, which would inevitably accompany the IC radiation,
substantially violates the current upper limits
at radio frequencies (Fig.~\ref{fig:model}b). To reconcile this discrepancy would require a magnetic field strength, $<$0.05 mG, substantially lower than current estimates for 
the GC \citep[e.g.,][]{Croc10}.
Furthermore, the IC cooling timescale is estimated to be $\sim$$5\times10^{5}{\rm~yr}$, which is inconsistent with the apparent length of G359.944-0.052 for any reasonable jet velocity.
We have also examined models involving synchrotron self-Compton and reached similar conclusions.






\section{Discussion and summary} {\label{sec:discussion}}
The above analysis lends support to our understanding of the striking spatial relationship among Sgr A*, the shock front and the linear X-ray feature G359.944-0.052:
a jet emanating from Sgr A* collides with and drives a shock front into the Eastern Arm, resulting in enhanced thermal emission in the radio and infrared bands.
The shock also accelerates a population of ultrarelativistic electrons that are responsible for the observed X-ray emission, most likely synchrotron,
as these electrons stream down the jet path. Before further addressing the physical properties of this jet, some remarks on the possible alternative nature of 
G359.944-0.052 are warranted. 

\subsection{Is G359.944-0.052 a pulsar wind nebula?}
\cite{Muno08} catalogued 34 extended X-ray features in the central 10-arcmin of the GC and suggested twenty of them
to be pulsar wind nebula (PWN) candidates (see also \citealt{Lu08,John09}), because the sizes, luminosities and power-law spectra of these candidates 
resemble the X-ray properties of known Galactic PWNe, in particular those with a bowshock-tail structure thought to be produced by a supersonically moving pulsar (see reviews by \citealt{Gaen06,Karg08}). 
G359.944-0.052 shows a power-law photon-index consistent with the typical values of PWNe ($1{\lesssim}{\Gamma}\lesssim2$), 
and its luminosity falls close to that of the least luminous PWNe known (\citealt{Karg08}).
Can G359.944-0.052 thus be the trace of a pulsar travelling supersonically in the GC? 

We note that the morphology of G359.944-0.052 makes it uniquely distinct from the PWN candidates and other extended X-ray 
features in the GC. As shown in \cite{Muno08} and confirmed by our deeper {\it Chandra} image, 
the PWN candidates, albeit in general characterized by an elongated morphology, are all resolved in both dimensions and almost always exhibit a bent or curved morphology, which, in the PWN scenario,
can arise from the bowshock-tail structure further subject to perturbation by the ambient medium (e.g, \citealt{Wang02,Lu03}).
In contrast, G359.944-0.052 is unresolved along its short-axis and appears straight and highly uniform along its long-axis. No obvious ``head'' can be identified for G359.944-0.052, 
which would be expected from a wind termination shock around a putative pulsar \citep{Gaen06}; the apparent ``knot'' near the middle of G359.944-0.052, if not an interloper, 
accounts for only $\lesssim$10\% of the total flux, and that it appears softer than its vicinity (Fig.~\ref{fig:xray}b) also argues against it being the site of the termination shock.
To our knowledge, only two pulsars, PSR J0357+3205 \citep{DeLu11} and PSR B2224+65 (the Guitar nebula; \citealt{John10}), are seen to exhibit a parsec-long, needle-shaped X-ray tail, 
the nature of which remains to be understood. In both cases, the X-ray flux of the pulsar itself amounts to 15\%-30\% of that of the tail, which
is typical of PWN systems ($\sim$25\%; \citealt{Karg08}) but markedly different from G359.944-0.052.
Moreover, numerical models suggest that a PWN can harbor a long and narrow tail only when it has a Mach number $M=V_{\rm p}/c_{\rm s}\gg$1 \citep{Gaen06}, where $V_{\rm p}$ is the pulsar velocity and $c_{\rm s}$
the sound speed of the ambient medium. As already pointed out by \cite{Muno08}, it is not easily conceivable how a pulsar can achieve a high Mach number in the GC, which is thought to 
be filled with a hot gas at a temperature $\gtrsim$1 keV and hence with $c_{\rm s}\gtrsim500{\rm~km~s^{-1}}$ \citep{Baga03}. A pulsar mimicking G359.944-0.052 might 
happen to run into a dense, cool cloud, but it would then require some fine-tuning to explain its needle shape, given the highly turbulent nature of the inner parsec of the GC.
In fact, the only PWN candidate seen within the inner parsec, G359.95-0.04, shows a substantial short-axis-to-long-axis ratio ($\sim$0.3), a prominent head and an appreciably curved tail \citep{Wang06}.
Given the above considerations, we strongly disfavor the PWN scenario for G359.944-0.052.

Indeed, \cite{Muno08} originally considered G359.944-0.052 as a possible jet from Sgr A* rather than a PWN. The geometric relationship among G359.944-0.052, Sgr A*, and the probable shock front in the Eastern Arm 
further underscores the unusual nature of G359.944-0.052. If it were a PWN, the spectral softening on the far-side (with respect to Sgr A*) of G359.944-0.052 would 
require that the putative pulsar move toward, rather than away from, Sgr A*. Such a motion, however, would be contrary to the momentum responsible for the shock 
front (i.e., arriving from the west; \S~\ref{sec:result}), or it would leave us with the conclusion that the putative pulsar is not related to the shock front, their remarkable
alignment is a mere coincidence, and the cause of the shock front remains unidentified. Instead, as we have discussed above, the hypothesis of a jet from Sgr A* offers a natural
explanation for a range of multi-wavelength phenomena. A number of interesting implications of the jet properties are in order.

\subsection{Inferred jet properties}
First, we surmise that the jet axis is aligned with the Galaxy's rotation axis (i.e., inclined by an angle of $\sim$90$^\circ$
to the line-of-sight); it is unlikely that a randomly tilted jet would be projected as close to the Galaxy's rotation axis as observed. This is further supported by the inferred location of the shock front, which, according
to the dynamical modeling of \citet{Zhao09}, lies approximately where the Eastern Arm passes through the Galaxy's rotation axis. In principle, the jet axis
should reflect the angular momentum of either the instantaneous accretion flow or the SMBH \citep{Blan77,Blan82,Meie01}. It is widely accepted that current accretion onto Sgr A* is mainly
supplied by winds from the circumnuclear stars \citep[e.g.,][]{Cuad08}, which are unlikely to collectively ``feel'' the global Galactic rotation due to their
short-lived nature. Thus it can be conjectured that the jet orientation is governed by the spin of the SMBH, which we suggest
has been coupled to the Galaxy's mean angular momentum via accretion over a Hubble time, if there have been no recent major merger events.
Our inferred jet axis is in reasonable agreement with the latest jet models (Markoff et al.~2007), which favor
an inclination $\gtrsim$$75^\circ$ and a position angle between $\sim$$90^\circ$--120$^\circ$, and is also compatible with 
model-dependent constraints on the spin axis of Sgr A* derived in several recent works \citep{Meye07,Brod11}.
The fact that our inferred jet axis is distinct from those of the other previously suggested jet candidates, except perhaps for the low-frequency radio ridge \citep{Yuse86}, deserves some remarks.
Indeed, the short dynamical timescale associated with the putative parsec-scale jet likely excludes the co-existence of other jets in the circumnuclear volume, in particular, that manifested 
by the chain of radio features roughly along a position angle of 60$^\circ$ (\citealt{Yuse12}; we offer a critique of this jet hypothesis in the Appendix).
On the other hand, we consider it premature to exclude the co-existence of large-scale collimated outflows that might operate on longer timescales and at possibly different orientations if produced during periods of relatively high accretion rates. 
Interestingly, the {\sl Fermi bubbles} \citep{Su10} have been suggested to be created by bi-polar jets following the Galaxy's rotation axis \citep{Guo12}.

Second, it is noteworthy that the putative parsec-scale jet shows no sign of being two-sided, a generic characteristic often presumed by jet models.
The inferred $\sim$90$^\circ$ inclination precludes the effect of Doppler boosting being important.
In fact, it could have been impossible for us to infer the presence of the jet, had it not interacted with the local gas.
In this regard, the jet must not have come in contact with the part of the Northern Arm that lies in projection closer to Sgr A* than the shock front on the Eastern Arm,
otherwise we might have found an additional shock. This conclusion is also supported by the dynamically inferred geometry of the Northern Arm \citep{Zhao09}.
Similarly, a counter-jet could exist without leaving an appreciable imprint, at least within the inner parsec, since it would lie
behind the Western Arc \citep{Zhao09}, and further out, be hidden by the dense gas of the {\it circumnuclear disk} \citep{Gust87}.
Regardless of the existence of a counter-jet, the shape and extent of the shock front strongly suggest
the presence of a secondary outflowing component of lower momentum, which might be understood as a sheath or a cocoon surrounding the jet \citep{Bege84},
with an opening angle of $\sim$$25^\circ$ (Fig.~2a). This estimate should be treated with caution since it is limited by the apparent width of the Eastern Arm.

Next, we consider possible constraints on the jet/outflow energetics and its potential impact on the GC environment.
A simple constraint comes from the input electron power in the above synchrotron models for G359.944-0.052, which spans a substantial range between $7\times10^{35}$ and $5\times10^{38}{\rm~erg~s^{-1}}$.
Alternatively, assuming that the apparent gas depletion and velocity discontinuity in
the immediate western side of the shock front (\S~\ref{sec:result}) are due to ram pressure imposed by the jet,
we may equate the jet kinetic power to the rate of change in the mechanical energy of the gas flow passing the shock front, $[GM_{\rm BH}{\mu}m_{\rm H}n_f/R][\pi(R\theta/2)^2] v_f \approx 
2\times10^{37}(R/0.5{\rm~pc})(\theta/25^\circ)^2(n_f/10^{4}{\rm~cm^{-3}})(v_f/200{\rm~km~s^{-1}}){\rm~erg~s^{-1}}$. Here $M_{\rm BH}$ is the mass of Sgr A*, $R$ the distance of the 
shock front from Sgr A*, $\theta$ the jet opening angle, and $n_f$ and $v_f$ are the density and Keplerian orbital velocity of the gas flow in the Eastern Arm, respectively \citep{Zhao09}. 
For comparison, the total available accretion power (in terms of the rate of accreted rest mass energy) 
of Sgr A* is $\sim$$10^{41}{\rm~erg~s^{-1}}$ near its Bondi radius (Cuadra et al.~2008) and drops by a factor of $\sim$100 near 10 Schwarzschild radii, 
because only a tiny fraction of the matter passing through the Bondi radius can sink to such a depth \citep[e.g.,][]{Yuan12}, 
where the jet is generally thought to be launched; an even smaller fraction of the accreted matter goes into the jet, the kinetic power of which is $\sim$$10^{37}-10^{38}{\rm~erg~s^{-1}}$ according 
to recent numerical simulations \citep{Yuan12b}.
On the other hand, the radiative luminosity of G359.944-0.052 predicted by our synchrotron model is only $\sim$$3\times10^{34}{\rm~erg~s^{-1}}$,
and from a dereddened Paschen-$\alpha$ luminosity of $\sim$$2.3\times10^{30}{\rm~erg~s^{-1}}$ from the shock front \citep{Wang10} we estimate
its total radiative cooling rate to be on the order of $\sim$$10^{33}{\rm~erg~s^{-1}}$.
Thus the radiative dissipation accounts for only a tiny fraction of the estimated jet power.
This is consistent with the sharp and straight appearance of G359.944-0.052, which indicates that the jet remains highly collimated after punching through the Eastern Arm. 
A better accounting of the jet energetics and kinematics can in principle be obtained by a self-consistent modeling of both the jet-Eastern Arm interaction and the 
broadband spectral energy distribution of Sgr A*, which is beyond the scope of this work. 
While it can be expected that the jet power will ultimately be consumed by the ISM beyond the inner parsec, the details of this process 
would largely depend on the external (thermal plus magnetic) pressure distribution, which is currently quite uncertain \cite[e.g.,][]{Guo12}.
In any case, the above inferred jet power is small compared to the collective kinetic power supplied by the winds of the circumnuclear stars ($\sim$$10^{39}{\rm~erg~s^{-1}}$; 
\citealt{Mart07}). This complicates the identification of possible interplay between the jet and the GC environment. 
For instance, the diffuse bipolar X-ray structure seen on a scale of $\sim$20 pc roughly along the Galaxy's rotation axis is more likely powered by the
circumnuclear stars rather than by the current level of jet activity \citep{Baga03,Morr03,Hear13}.

The above discussion rests on the implicit assumption of a steady jet. The limited counting statistics and timespan of the X-ray data do not allow 
us to assess short- or long-term variations in the jet energetics. The former might be related to the observed flares from Sgr A*, which have been
suggested to be generated by, among other possibilities, ``episodic jets'' that can release energy two orders of magnitude higher than in the quiescent state \citep{Yuan09}. 
Even stronger activity of Sgr A* in the recent past ($\gtrsim$100 yrs ago) has been inferred from the 6.4-keV Fe-K$\alpha$ line reverberation in molecular clouds in the GC (e.g., Morris et al.~2012; \citealt{Pont13}). 
Significantly greater jet energetics in the past are also suggested by the 30-pc low-frequency radio ridge, which could arise from relatively low-energy electrons in the far downstream portion of our proposed jet. 
According to \cite{Yuse86}, the monochromatic luminosity of this feature is $\sim$1000 times 
higher than that predicted by our favored synchrotron model at the position of G359.944-0.052 (Fig.~\ref{fig:model}a). 
Further sensitive radio observations are warranted in order to determine whether G359.944-0.052 can be more closely linked to the 30-pc ridge.
Finally, we expect an ultimate test of the suggested jet orientation by VLBI observations toward the event horizon of Sgr A* in the near future.




\acknowledgments
We wish to express our gratitude to Jun-Hui Zhao for providing us with the VLA images, and to John Lacy for 
his assistance with the TEXES data cube. We thank Sera Markoff, Gunther Witzel and Feng Yuan for helpful discussions.
MRM was partially supported by NSF grant AST-0909218.
FKB was supported in part by SAO grant GO3-14099X under NAS8-03060.






\appendix
\section{Commentary on the 3-parsec-scale radio jet candidate}
As noted in \S~\ref{sec:discussion}, the short dynamical timescale makes it unlikely that there are multiple parsec-scale jets from Sgr A* occupying the same radial range, but having very different orientations.
Recently, Yusef-Zadeh et al.~(2012, hereafter YZ12) proposed the presence of a jet from Sgr A* within the inner 3 pc of the Galactic Center (GC). The backbone of that proposal is an apparent linear striation in radio images (in particular, at 22 GHz) passing through Sgr A* and oriented at a position angle (PA) of $\sim$$60^\circ$ (east from north), accompanied by a chain of compact radio-emitting features roughly along the same line and roughly colinear with Sgr A*.  Because this hypothesized jet has a very different orientation from the one that is discussed in this paper, and because both hypotheses are unlikely to be simultaneously valid, we have closely examined the YZ12 hypothesis.  In the arguments detailed in this appendix, we offer some of the reasons why we disagree with the assertion by YZ12 that the radio features trace a coherent, linear outflow structure, and we therefore question the evidence for their proposed jet. 

\begin{figure}
\epsscale{1.0}
\plotone{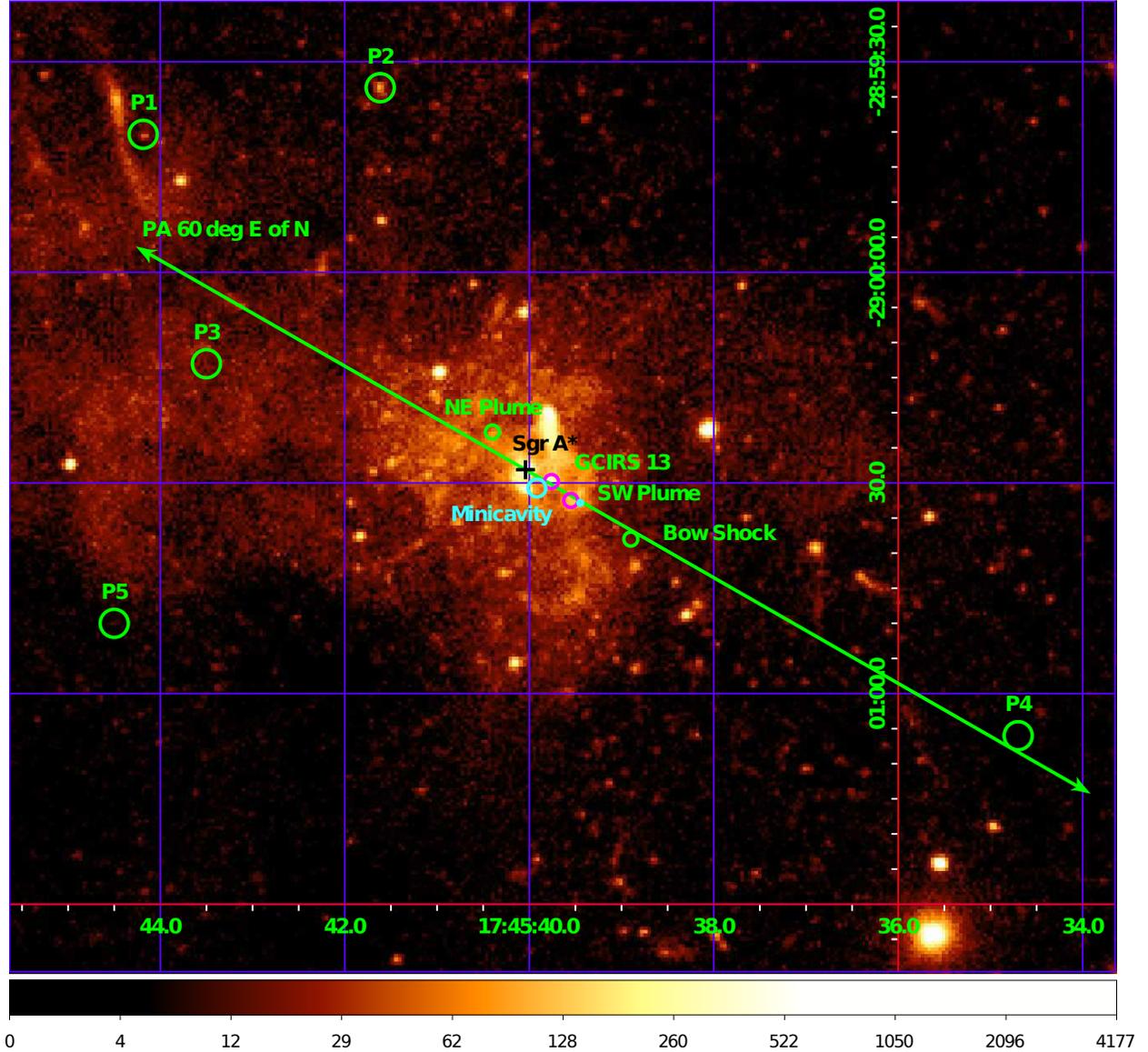}
\caption{A {\it Chandra}/ACIS 2-8 keV image of the GC.
The positions of various radio and X-ray features discussed in YZ12 are marked by circles. The green line represents the proposed radio jet along PA of $60^\circ$ ($240^\circ$).
Note that both P1 and P2 seem to have a compact, rather than extended, X-ray counterpart, as cataloged by Muno et al.~(2009), at $[RA, DEC]=[266.43407,-28.99466]$ and [266.42344, -28.99278], respectively. 
P4, showing no X-ray counterpart, lies approximately at $[RA, DEC]=[266.39583,-29.018056]$ and is not colinear with P1 and Sgr A*. Celestial coordinates (J2000) are shown.}
\label{fig:a1}
\end{figure}

\begin{figure}
\epsscale{1.0}
\plotone{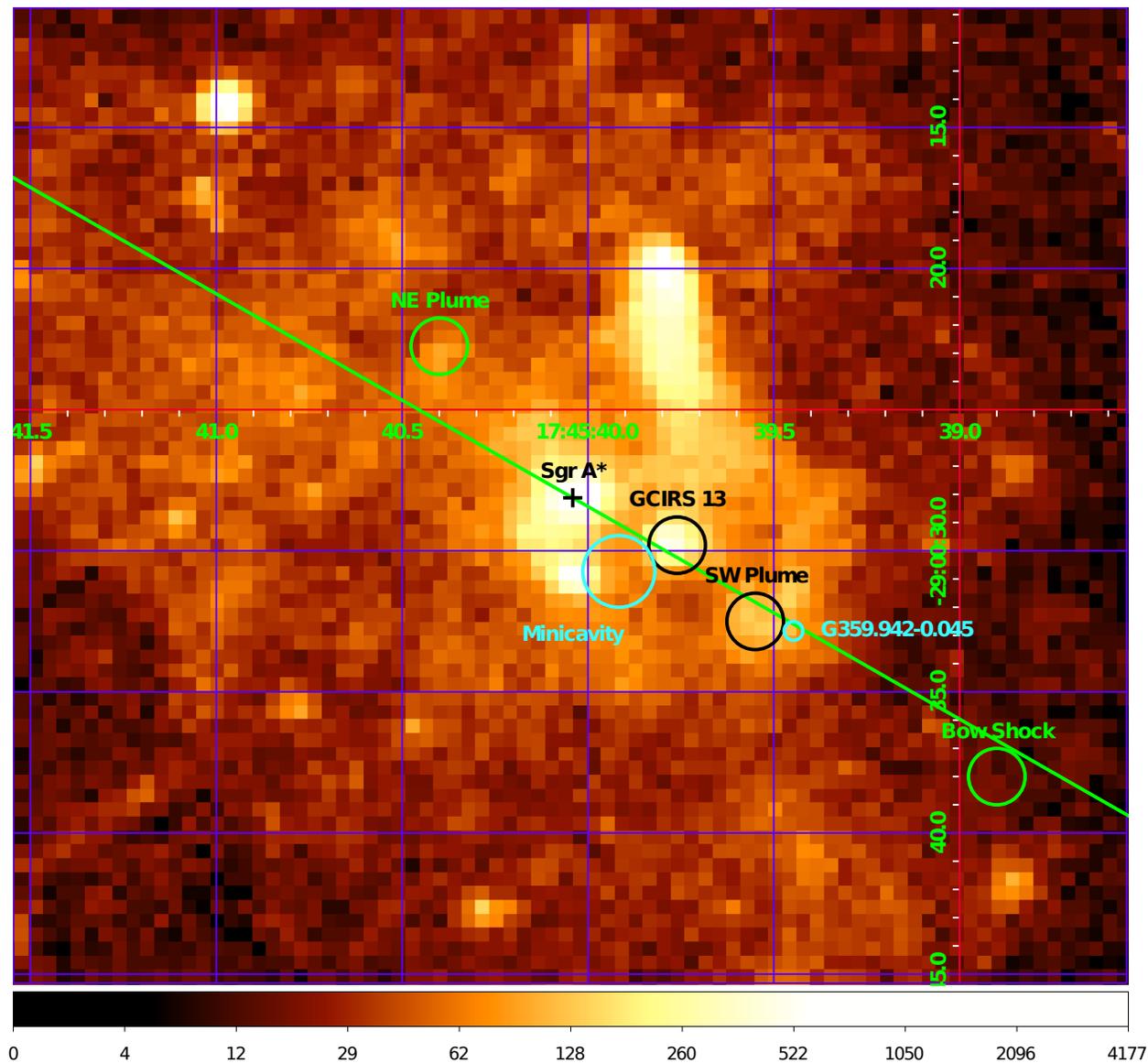}
\caption{Similar to Fig.~\ref{fig:a1} but for a close-up view of the vicinity of Sgr A*. The NE Plume lies on an extended ridge of X-ray emission running SE to NW, while the SW Plume connects to extended
X-ray emission running to the NW. There is no significant X-ray emission associated with the putative ``bow shock'' noted by YZ12.}
\label{fig:a2}
\end{figure}

(i) YZ12 reported the finding of several radio maxima (blobs) within one parsec of Sgr A*, which are distributed along ``a faint, continuous linear structure centered on Sgr A* with a PA $\sim$$60^\circ$'' 
(see their Figures 1 and 4), as might naturally be expected if they had been produced by a highly collimated jet. However, as is evident from their figures, and from Figs.~\ref{fig:a1} and \ref{fig:a2} here, 
these radio blobs are only very loosely distributed along a straight line passing through Sgr A*.  
They rather appear as an incoherent set of local enhancements distributed about the line representing the putative jet.  
There are a number of separate complex structures in the central parsec with which the blobs can be identified.  In particular, blobs $d$ and $e$ are located on the Western Arm of the Sgr A West HII region; 
they also appear in the 22 and 8.4 GHz radio continuum images presented by Zhao et al.~(2009; named X21 and X23 therein). The proper motions of these two blobs measured by Zhao et al.~(2009; their Figure 12c)
show no significant velocity component along the proposed jet path (i.e., near PA-240$^\circ$), as might be expected if they are genuine sites of interaction between a jet and the Western Arm. 
Instead, Zhao et al.~(2009) modeled the observed motions satisfactorily in terms of coherent rotational motions of the inside edge of the circumnuclear disk about Sgr A*, such that the bulk
motion is roughly perpendicular to the proposed jet path.

(ii) YZ12 noted several resolved features of 8.4 GHz polarized emission\footnote{Unfortunately, the calibration and reliability of these features were not discussed.} beyond Sgr A West (named P1-P5; see their Figure 3). In particular, they suggested that P1 and P4, placed 
roughly symmetrically about Sgr A*, are the apparent endpoints of the aforementioned linear structure.  We show in Fig.~\ref{fig:a1} the relative positions\footnote{These positions are extracted from the figures in YZ12, 
as the positions listed in their text are apparently erroneous.} of the various features that YZ12 propose as possible tracers of a jet, superimposed on our {\it Chandra}/ACIS image of the GC.  
As can be seen, sources P1 and P4 are {\it not} colinear with Sgr A*. P1 and P4 also differ in that P1 has a compact X-ray counterpart while P4 does not (Fig.~\ref{fig:a1}). 
Furthermore, sources P1 and P3 are near the center of the well-defined supernova remnant (SNR), Sgr A East, so the existence of nonthermal radio emission in that region has a natural explanation unrelated to a jet.  
Finally, sources P2 and P3 coincide with ridges of HCN emission from the circumnuclear disk (Christopher et al. 2005), so interaction of that disk with the rapidly expanding SNR offers an alternative possibility for the production of those polarized features.

(iii) YZ12 suggest two extended X-ray-emitting features, which they denoted ``NE Plume'' and ``SW Plume'' (Fig.~\ref{fig:a2}), as sites of interaction between the jet and the gas in the arms of 
Sgr A West.  The ``NE Plume'' has previously been reasonably interpreted as part of the semi-circular X-ray arc generated by the interaction of the interior ejecta of 
the Sgr A East SNR with the stellar winds of the central parsec cluster (Rockefeller et al. 2005).  Moreover, the ``NE Plume'' is substantially offset from PA-60$^\circ$ 
(YZ12's schematic picture -- their Figure 5c -- is quite misleading in this regard). The ``SE Plume'' likewise appears to connect to extended X-ray emission running to the NW.  
In our opinion, these extended structures argue in favor a broad-angle stellar-wind-driven outflow rather than a collimated jet. 
Indeed, a high mass-loss-rate star, AF-NW, is present at that location (see Figure 5 in \citealt{Baga03}).

(iv) While evidence also exists for a wind/outflow coming from the general vicinity of Sgr A* toward the mini-cavity, based on the 
cometary shape of two IR sources \citep{Muz07,Muz10}, the origin of this outflow remains uncertain. Indeed, it has been suggested and modeled that the outflow (and hence the {\it mini-cavity} and the radio blobs) 
is produced by stellar winds from the central cluster of massive, young stars (the IRS16 cluster), perhaps collimated by the gravitational 
potential of Sgr A* (Wardle \& Yusef-Zadeh 1992; Melia, Coker \& Yusef-Zadeh 1996). In this regard, a jet or a collimated outflow from Sgr A* is not the only viable interpretation for those features.  

The current paradigm is that black holes endowed with spin and accreting magnetized plasma will produce an outflow, although the degree of collimation and the power of the outflow may depend on several parameters.  Indeed, the accretion rate onto the Galactic black hole is so much smaller than that which is apparently flowing through its Bondi radius that most of the matter must 
be leaving in an outflow of some kind (e.g., Genzel et al. 2010).  Finding the strength, orientation, and degree of collimation of this outflow is clearly an important endeavor.  The YZ12 hypothesis must be considered as an interesting possibility, but we suggest that the evidence for it needs to be strengthened before it be accepted as a competitive hypothesis for a jet.

\clearpage


\clearpage


\end{document}